\documentclass[onecolumn,showpacs,preprintnumbers,amsmath,amssymb,nofootinbib]{revtex4}
\usepackage{graphicx}
\usepackage{bm}
\usepackage{latexsym}
\usepackage{epsfig}
\usepackage{natbib}
\usepackage{color}

\begin{document}

\title{Decaying Vacuum Inflationary Cosmologies: A Complete Scenario Including Curvature Effects}

\author{J. A. S. Lima\footnote{jas.lima@iag.usp.br}}

\address{Departamento de Astronomia, Universidade de S\~ao Paulo, Rua do Mat\~ao 1226, 05508-900, S\~ao Paulo, Brazil\\
limajas@astro.iag.usp.br}

\author{E. L. D. Perico}

\address{Instituro de F\'isica, Universidade de S\~ao Paulo, Rua do Mat\~ao Travessa R 187, 05508-090, S\~ao Paulo, Brazil\\
elduartep@usp.br}

\author{G. J. M. Zilioti}

\address{Departamento de Astronomia, Universidade de S\~ao Paulo, Rua do Mat\~ao 1226, 05508-900, S\~ao Paulo, Brazil\\
gzilioti@usp.br}

\date{\today }

\begin{abstract} We propose  a  large class of  nonsingular cosmologies of arbitrary spatial curvature  whose cosmic history is  determined by a primeval dynamical $\Lambda (t)$-term.  For all  values of the curvature, the models evolve between two extreme de Sitter phases driven by the relic time-varying vacuum energy density. The transition from inflation to the radiation phase is universal and points to  a natural solution of the graceful exit problem regardless of the values of the curvature parameter. The flat case recovers the scenario recently discussed in the literature (Perico et al., Phys. Rev. D88,  063531, 2013). The early  de Sitter phase is characterized by an arbitrary energy scale $H_I$ associated to the  primeval vacuum energy density. If $H_I$ is fixed to be nearly the Planck scale, the ratio between the relic and the present observed vacuum energy density is $\rho_{vI}/\rho_{v0} \simeq 10^{123}$.
\end{abstract}

\maketitle
\section{Introduction}

Probably, the most beautiful developments of modern observational cosmology are related to the discovery of the current accelerating stage of the universe and the evidence about primordial inflation. Both of them require the presence of new agents on the cosmic inventory different from  radiation,  baryons and cold dark matter (CDM).  

In the framework of general relativity, it is widely believed that some sort of dark energy  provides the fuel driving  the  recent accelerating stage of the universe whose top ten candidate is represented by the cosmological $\Lambda$ - term. After more than one decade of detailed studies, it was shown that the so-called flat $\Lambda$CDM model  fits very well the current high quality cosmological data as provided by the  supernovae type Ia (SN Ia) observations\cite{Riess98,Perlmuter98,AmanullahETAJ10}, anisotropies of the cosmic microwave background radiation (CMB)\cite{KomatsuETAJ11,Planck14},  baryon acoustic oscillations (BAOs)\cite{BeutlerETALMNRAS11,BlakeETALMNRAS11,AndersonETALMNRAS12} and complementary observations (galaxy clustering\cite{BPL10,P13,M10}, weak lensing\cite{WL13},  H(z) parameter\cite{Ratra13}, growth factor\cite{basilakos}, etc.)

Despite the great success of the cosmic concordance $\Lambda$CDM model, the presence of a cosmological constant (vacuum energy density, $\rho_v \equiv \Lambda /8 \pi G$) gives rise to a rather unsettled situation in the interface uniting particle physics and cosmology, sometimes dubbed the cosmological constant problem (PCC). Briefly, it is puzzling that the present day upper bound of $\Lambda$ (as provided by measurements of the current Hubble parameter) differs from natural theoretical expectations by more than 100 orders of magnitude\cite{WeinbergRMP89,PeeblesRatraRMP03}. 
Interestingly, long  before the SN Ia observations establishing the current accelerating stage of the Universe, the PCC together the age of the Universe problem inspired many authors to propose cosmologies driven by vacuum energy density that decays continuously  in the course of the evolution\cite{OzerTahaPLB86,OzerTahaNPB87,FreeseETNPB87,ChenWuPRD90,PavonPRD91,CarvalhoETPRD92,Waga93,LimaMaiaPRD94, SpindelB94,ArbabAbdelPRD94,LimaTroddenPRD96,BloomfieldWagaMNRAS96,LimaPRD96}(see also the  paper by Overduin and Coperstock\cite{OverduinCooperstockPRD98}, and Lima\cite{Lima2004} for  short reviews on this old literature). 

The continuous decay process happens because the vacuum is coupled with the other matter fields, and, as such, these models can be seen as the legitimate predecessors of the  interacting dark matter-dark energy cosmologies\cite{WPCPLB2001,ERPRD2009,Fabris10}. In order to alliviate the PCC, the huge decaying vacuum energy density  must be responsible by the early  inflation (de Sitter phase), while  at the late stages provides the expected ratio between its primordial and current day value\cite{LimaMaiaPRD94,LimaTroddenPRD96}. Dynamical vacuum models have also been suggested to solve the so-called coincidence problem\cite{Stein97,AL05}, that is,  the fact that the time varying matter-energy density and the (constant) vacuum energy density have the same order of magnitude nowadays.    

Recently, a flat decaying $\Lambda$(H)-model describing  both the early inflation and the current accelerating stage as described by the standard $\Lambda$CDM model was proposed in the literature\cite{LBS13,PLBS13}. It was shown that models  of this kind are easily implemented by using a $\Lambda(H)$-term that at early times grows faster than $H^2$.  The primordial de Sitter phase is unstable and the model is driven continuously to the standard radiation phase. Subsequently, it evolves to the $\Lambda$CDM  description driven by the leftover vacuum energy density. In other words, the model accommodates the eras described by the cosmic concordance model plus inflation. However,  unlike many inflationary variants endowed with a preadiabatic phase, the nonsingular de Sitter stage is nonadiabatic and progressively generates the now observed radiation entropy content\cite{LBS14} within our horizon ($S_0 \sim 10^{88}$ in natural units). This means that the extreme adiabatic supercooling and the dramatic reheating process of scalar field inflationary models is absent and the horizon, the ``graceful exit" and the coincidence problems have a natural explanation within this complete scenario. 

On the other hand, a flat de Sitter spacetime  as assumed  in the quoted papers\cite{LBS13,PLBS13,LBS14} is past-geodesically incomplete since it covers only a half of the compact de Sitter hyperboloid. In this concern,  a  closed de Sitter primordial space  is endowed with some nice theoretical features, among them: (i) it is in agreement with the old idea of a  universe ``created from nothing'' as discussed by many authors\cite{Zeld,Vilenkin82} (ii) it covers the whole de Sitter hyperboloid, and, as such, it is geodesically complete. 

Further, compact closed models explains  naturally why the net charge  of the observed cosmic structures is incredibly adjusted to zero. Indeed, for every closed surface located in a compact space the sum of the charges on the two sides of the surface is zero\cite{Landau1985}.  Spatially compact closed model  also predict that the total momentum and energy is zero, a well known  result also in line  with the idea of ``creation from nothing".  In this way, a remnant closed $\Lambda$CDM with a very small curvature nowadays, resembling  a `quasi'-flat de Sitter space (after many aeons of tepid inflation and decelerating expansion), may also describe the current observed Universe. In a point of fact, although considering that the flat case is favored by the present data, the possibility of curvature still deserves a closer scrutiny  because the Planck results combined with WMAP yield for the curvature density parameter\cite{Planck14}, $\Omega_{\kappa}= -0.037^{+0.043}_{-0.049}$ ($95\% c.l.$).  Therefore, since the error bars are still slightly compatible with the closed and hyperbolic cases, it is interesting to investigate what happens when  the flat condition is relaxed in a more general treatment. 

In this context, we investigate here if the primeval accelerating de Sitter stage can be generated by a decaying $\Lambda$-term regardless of the sign of the spatial curvature parameter. The main aim is to understand how this kind of complete cosmological history (``from de Sitter to de Sitter") depends on the curvature parameter.

The article is organized as follows. In section 2, we set up the basic equations for an arbitrary decaying vacuum energy density model. In section 3, we review briefly some properties of the de Sitter spacetime with emphasis on the comparison between the closed and flat cases thereby reinforcing the search for a more general framework. In section 4, we consider a phenomenological approach to obtain the general $\Lambda(t)$-term including curvature adopted in our work. The dynamic properties of the complete scenario are discussed in section 5, and, finally, the main conclusions are summarized in section 6.

\section{Decaying-$\Lambda$ Models: Basic Equations}

In this work we shall consider spatially homogeneous and isotropic spacetimes described by the FLRW metric:

\begin{equation}\label{metric}
ds^2=dt^2-a(t)^2\left[{d\chi^2} + \sigma^{2}(\chi)(d\theta^2 + \sin^{2}\theta d\phi^{2})\right]\,,
\end{equation}
where $a(t)$ is the scale factor,  $\kappa=\pm 1, 0$ is the normalized spatial curvature parameter,  and the function $\sigma(\chi)$ is given by:

\begin{equation}\label{sigma}
\sigma(\chi) = \frac{\sin\sqrt \kappa\,\chi}{\sqrt \kappa} = \left\{\sin\chi,\, \sinh\chi,\,\chi \, \Leftrightarrow \, \kappa =\pm 1, 0 \right\}.
\end{equation}
where $\kappa$ is the normalized spatial curvature parameter and the $\chi$-coordinate takes values on the intervals: $0 \leq \chi \leq \pi$, $0 \leq \chi \leq \infty$,  for $\kappa\pm 1,\,0$, respectively.  

In such a background the Einstein field equations for an interacting two-fluid mixture (vacuum plus a perfect simple fluid) can be written as:

\begin{equation}\label{rho_T}
8\pi G\rho_T \equiv 8\pi G (\rho + \rho_v) = 3\frac{\dot a^2}{a^2} + 3\frac{\kappa}{a^2}\,,
\end{equation}
and

\begin{equation}\label{P_T}
8\pi G p_T \equiv 8 \pi G(p - \rho_v) = -2\frac{\ddot a}{a}- \frac{\dot a^2}{a^2} - \frac{\kappa}{a^2}\,,
\end{equation}
where $\rho$, p are the energy density and pressure of the fluid component, $\rho_v \equiv \Lambda(t)/8 \pi G$ is the vacuum energy density, and an overdot means time derivative. 
The Bianchi identities imply that the total energy density is conserved: 
\begin{equation}
\dot\rho_T+3H(\rho_T+P_T)=0\,,  
\end{equation}
where $H\equiv\dot a/a$ is the Hubble expansion rate. In terms of the fluid components the above expression  takes the form:
\begin{equation}
\dot\rho+ 3H(\rho + p)= -\dot\rho_v\,. 
\end{equation}
This equation shows that a decaying vacuum energy density or equivalently, a decaying $\Lambda(t)$ model is possible only if the vacuum medium changes energy with the fluid component. As shown long ago, if the vacuum is interpreted as a kind of condensate which does not carry entropy, the early $\Lambda (t)$-term is time decreasing function\cite{LimaPRD96} so that the vacuum energy density is small today because the Universe is too old. 

In what follows, we are particularly interested in the solutions of the above equations capable to generate a complete cosmic history as suggested by inflation and the late time observations regardless the value of the curvature parameter.  This means that the spacetime must start and finish with two extreme accelerating de Sitter phases intercalated by two decelerating  eras  (radiation and nonrelativistic matter phases). 

\section{Some Properties of the de Sitter Spacetimes}
  
The so-called de Sitter models are the set of solutions satisfying the equation of state $P_T = -\rho_T$. These models can physically be realized by two different manner: i) a pure cosmological constant (vacuum energy density) with $\rho=p=0$, and (ii) an exotic simple fluid obeying the equations of state $p=-\rho$ ($\Lambda=0$). 

From a geometrical viewpoint such spacetimes are characterized  by a constant and positive scalar of curvature $R \equiv g_{\mu \nu} R^{\mu\nu}$ with the Riemann tensor determined only by this Ricci scalar\footnote{$R^{\mu\nu\alpha\beta} = \frac{R}{12}(g^{\mu \alpha}g^{\nu \beta}-g^{\mu \beta}g^{\nu \alpha})$}. Like the Minkowski flat geometry, de Sitter spacetimes also possess a ten-parameter group of symmetry or group of motions (10 Killing vectors), and the current interest for such models comes from the fact that both the very early stages, as well as the future universe can be described by this kind of cosmology. 

Henceforth it will assumed that the  pressure of the ordinary perfect fluid is always positive, and, as such,  the possible de Sitter solutions are always generated when the energetic content is for all practical purposes represented by the dominant $\Lambda$-term. For $P_T = -\rho_T$, we see from (2)-(3) that the differential equation driving the evolution of the scale factor is given by 
\begin{equation}\label{EqSitter}
a\ddot{a} - \dot{a}^2 - \kappa = 0,
\end{equation} 
whose solutions can be written as\cite{LimaTroddenPRD96} 

\begin{equation}\label{desitter}
a(t)=\left\{
\begin{array}{lll}
H_I^{-1}\text{cosh}(H_I t)\,, & \quad & \kappa=+1\\
a_\star	\exp(H_I t)\,, & \quad & \kappa=0\\
H_I^{-1}\text{sinh}(H_I t)\,, & \quad & \kappa=-1
\end{array}
\right.
\end{equation}
where $H_I = \sqrt{{\Lambda_I}/{3}}$. The subindex ``I" above refers to inflation (primordial de Sitter phase). Naturally, for the late time de Sitter era,  the parameter $H_I$ in the above expressions must be replaced by $H_F=\sqrt{\Lambda_{\infty}/3}$, the final value of the Hubble parameter. Note also that the Hubble parameter remains constant ($H(t)=H_I$) only in the flat case. As one may check, for $\kappa \pm 1$, $H_I$ is still a characteristic de Sitter energy scale but it specify the maximum (minimum) value of $H(t)$, respectively. 

On the other hand, it is widely known that de Sitter solutions as described  above can also be visualized as a 4-dimensional hyperboloid embedded in a flat 5-dimensional space-time\cite{S56,hawking,BD82}. In order to see that, it is convenient to write  the line element describing such spacetimes in the 5-dimensional form:

\begin{equation}
ds^2 = dz_0^2 - dz_1^2 - dz_2^2 - dz_3^2 - dz_4^2,
\end{equation}  
with the hyperboloid defined by the constraint:

\begin{equation}
z_0^2 - z_1^2 - z_2^2 - z_3^2 - z_4^2 = -\tau^2,
\end{equation}
where $\tau$ is a dimensional positive constant.

\begin{figure}[!ht]\centering
\includegraphics[width=8cm]{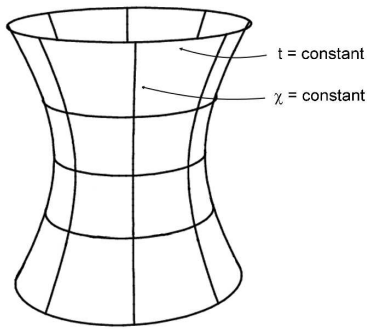}
\caption{The de Sitter manifold represented by a compact hyperboloid imbedded in a 5-dimensional flat space.  The circles are surfaces of constant $t$ and the vertical lines are normal geodesics which represent hypersurfaces of constant $\chi$ (increasing time coordinate).}
\end{figure}

Different parameterizations of this surface correspond to different values of the curvature. To show this for the closed case let us consider polar coordinates:

\begin{align}\label{coclosed}
z_0 &= \tau  \sinh\left(\frac{t}{\tau}\right), \nonumber \\
z_1 &= \tau  \cosh\left(\frac{t}{\tau}\right) \cos \chi, \nonumber \\
z_2 &= \tau  \cosh\left(\frac{t}{\tau}\right) \sin \chi \cos \theta, \\
z_3 &= \tau  \cosh\left(\frac{t}{\tau}\right) \sin \chi \sin \theta \cos \phi, \nonumber \\
z_4 &= \tau  \cosh\left(\frac{t}{\tau}\right) \sin \chi \sin \theta \sin \phi. \nonumber
\end{align}
As one may check, the hyperboloid equation is identically satisfied and the metric in this coordinate system assumes the following form:

\begin{equation}
ds^2 = dt^2 - \tau^2 \cosh^2\left(\frac{t}{\tau}\right) \left[d\chi^2 + \sin^2\chi \left(d\theta^2+\sin^2\theta d\phi^2\right)\right],
\end{equation}
which corresponds to the closed case ($\kappa = 1$)\footnote{The hyperbolic case ($\kappa = -1$) can be obtained by considering the hyperboloid in 5-dimensions under the prescription: $\sinh(t/\tau) \leftrightarrow i\cosh(t/\tau)$, $\chi \rightarrow i\chi$.}  as given by the first solution of \eqref{desitter} under the identification $\tau = \sqrt{\frac{3}{\Lambda}} \equiv H_I^{-1}$. The above metric is exactly the one given by \eqref{metric} with $\sigma(\chi) = sin\chi$ (see Eq. (\ref{sigma})).

Figure 1 shows the de Sitter hyperboloid with two dimensions suppressed.  The coordinate system introduced by \eqref{coclosed}  covers entirely the hyperboloid  and describes a closed de Sitter spacetime with the related coordinates ranging the standard values (compare with metric \eqref{metric}). Note also that the spatial sections of constant time shown in Fig. 1 are spheres $S^{3}$ and the normal geodesics (ortogonal lines) start from the distant past ($t= - \infty$) contract continuously to the smallest spatial separation and re-expand to infinity\cite{hawking}. 


Another possible set of coordinates adapted for the flat case ($\kappa=0$) is defined by:

\begin{align}\label{coflat}
u &= \tau \sinh \left(\frac{t}{\tau}\right) + \tau\frac{r^2}{2} e^{t/\tau}, \nonumber \\
v &= \tau \cosh \left(\frac{t}{\tau}\right) - \tau\frac{r^2}{2} e^{t/\tau}, \nonumber \\
a &= \tau x e^{t/\tau},  \\
b &= \tau y e^{t/\tau},\nonumber \\
c &= \tau z e^{t/\tau}, \nonumber 
\end{align}
with $r^2 = x^2 + y^2 + z^2$. The metric is

\begin{figure}[!ht]\centering
\includegraphics[width=8cm]{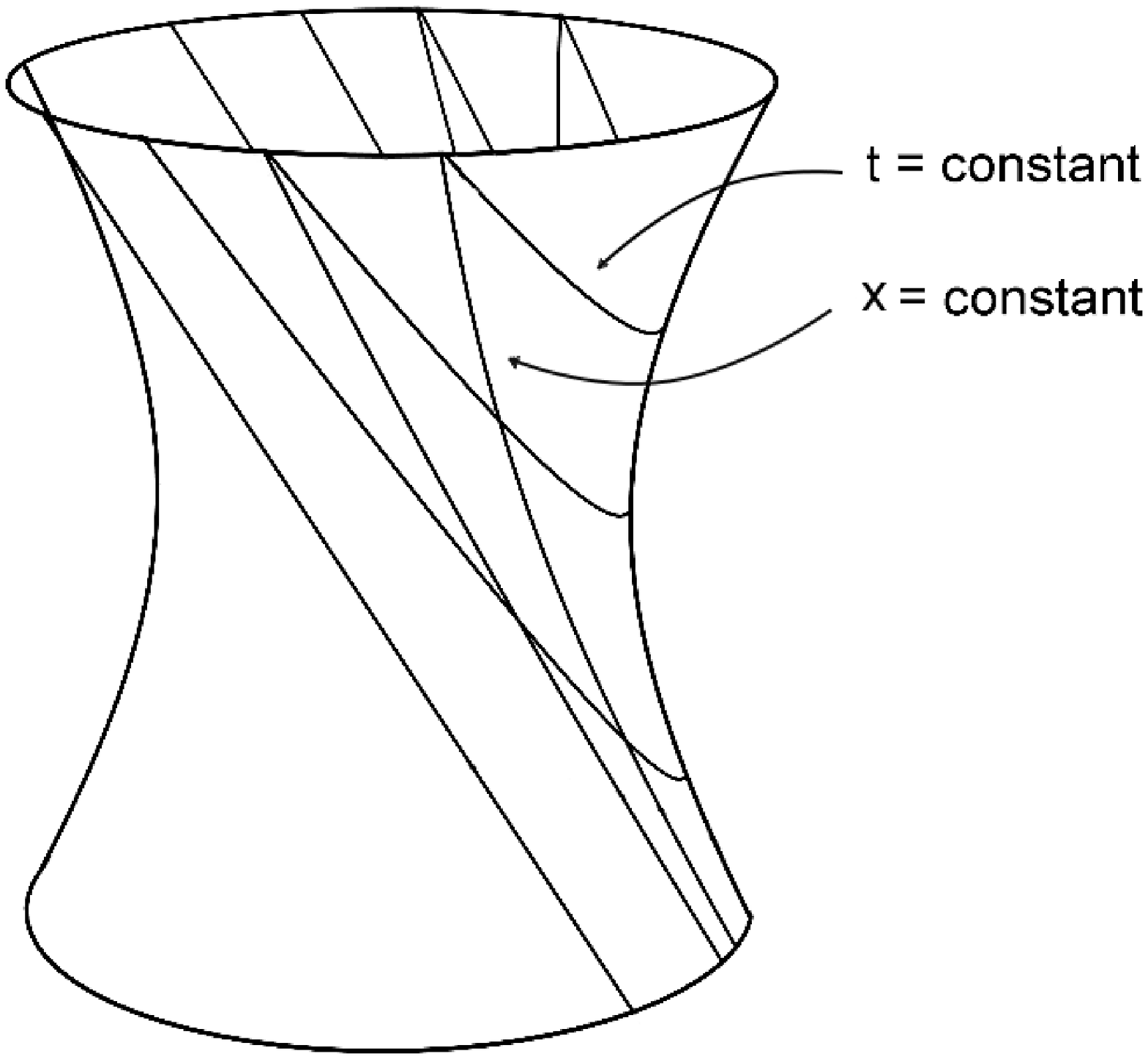}
\caption{\text{Coordinates in the flat case}. The same de Sitter hyperboloid of Figure 1 but now showing the coordinate system describing the flat case as given by Eqs. \eqref{coflat}-(\ref{flat}). Clearly it does not cover the compact 4-dimensional hypersurface (see text)}\label{fig:hyper} .
\end{figure}

\begin{equation}\label{flat}
ds^2 = dt^2  -\tau^2 e^{2t/\tau}\left[dx^2+dy^2+dz^2\right],
\end{equation}
which is just \eqref{desitter} for $\kappa = 0$ with $\tau = H_I^{-1}$ in Cartesian coordinates.  

Figure 2 illustrates the hyperboloid for this  set of coordinates (again for obvious reasons two dimensions have been suppressed). As it appears,  this choice of coordinates does not cover the whole hyperboloid since $u+v = \tau e^{t/\tau} \geq 0$ so that only half is covered. From this point of view, it seems interesting to investigate the generation of the early and late time de Sitter stages for $\kappa\, \pm \,1$ cases because unlike the flat de Sitter case such spacetimes are geodesically complete. 

\section{Decaying Vacuum Law Including Curvature}


The two-fluid mixture is assumed to satisfy the independent cosmological equations (\ref{rho_T})-(\ref{P_T}). In order to establish a model and study their
observational and theoretical predictions in $\Lambda(t)$ scenarios,  one needs first
to specify the time-dependence of $\Lambda(t)$. The existence of two energy components means that we may introduce phenomenologically the following ratio:

\begin{equation}
\beta \equiv \frac{\rho_v - \rho_{v0}}{\rho_T}\,,
\end{equation}
where $\rho_{v0}$ is a constant vacuum density associated to the very late time value, and $\rho_T=\rho+\rho_v$ is the total energy density, as defined by \eqref{rho_T}.
In principle, this $\beta$ parameter quantifies the time variation of the $\Lambda(t)$ term. It satisfies the following properties:

\begin{enumerate}
\item In the extreme limiting  case, $\rho_v \rightarrow \rho_{v0}$, we see that $\beta \rightarrow 0$ so that the model behaves like $\Lambda$CDM cosmology:
	\begin{equation*}
	8\pi G\rho_v =\Lambda=\Lambda_0\,.
	\end{equation*}
Naturally, at this limit the energy-momentum of the perfect-fluid component is identically conserved.
\item If $\rho_{v0}\ll\rho_v$, the ratio defines the fraction of the vacuum to the total energy density.
If this ratio remains constant during the periods of the cosmic history when the radiation or nonrelativistic matter are 
the dominant components we obtain\cite{CarvalhoETPRD92}:
	\begin{equation*}
	\Lambda(t)=3\beta\left(H^2+\frac{\kappa}{a^2}\right)\,.
	\end{equation*}
This means that the total vacuum contribution after inflation is  given by:
	\begin{equation*}
	\Lambda(t)=\Lambda_0 + 3\beta\left(H^2+\frac{\kappa}{a^2}\right)\,,
	\end{equation*}
It worth noticing that the limiting case $\kappa=0$ has been theoretically deduced based on renormalization group approach techniques\cite{ShapiroSolaPLB02}. 

\item Probably, the most impressive and natural possibility is that the ratio $\beta$ does not remain constant i.e. $\beta=\beta(t)$. In this case,  the vacuum energy density can be  generally expressed as:

	\begin{equation}
	\rho_v = \rho_{v0} + \beta(t) \, \rho_T,
	\end{equation}
	or equivalently, the $\Lambda(t)$ term can be written as:

	\begin{equation}\label{lambda}
	\Lambda(t) = \Lambda_0 + 3\beta(t)\left(H^2 + \frac{\kappa}{a^2}\right)\,.
	\end{equation}
\end{enumerate}

Now, for the sake of generality, we expand the dimensionless time-dependent $\beta$ parameter as a function of H(t):
	
\begin{equation}\label{ansatz}
\beta(t) = \nu + \alpha \left(\frac{H}{H_I}\right)^n\,,
\end{equation}
where $\nu$, $\alpha$  are dimensionless constants and $H_I$ is the Hubble parameter associated to the primordial de Sitter stages.  

By inserting the above expression  into (\ref{lambda}) and choosing $\Lambda_{\infty}$ as the value of the running $\Lambda (H,a)$ when $a \rightarrow \infty$,  $H=H_F \ll  H_I$ (where $H_F$ is the final constant value of H) we obtain:

\begin{equation}\label{LambdaG}
\Lambda (t) = \Lambda_{\infty}  + 3\nu\left(H^2 - H_F^{2} +\frac{\kappa}{a^2}\right)  + 3\alpha\left(\frac{H}{H_I} \right)^n \left(H^2+\frac{\kappa}{a^2}\right)\,.
\end{equation}
In the flat case, the above expression reduces to the one adopted by Perico et al.\cite{PLBS13} with a different notation (see their equation (15)). 

Although rather general, expression (\ref{LambdaG})  can be simplified based on different arguments. First, without loss of generality, we see that the parameter $\alpha$  can be absorbed in the value of the scale $H_I$ so that we may fix $\alpha =1$. Still more important, the parameter $\nu$  has been estimated using  theoretical arguments and current observational data.  In the flat case,  Basilakos, Plionis \& Sol\`a\cite{BPS09} obtained  $\nu \simeq 10^{-3}$  based on a joint analyses involving CMB, SNe Ia and BAO while a theoretical analysis by Sol\`a\cite{sola2008} yielded $|\nu| \sim 10^{-6}-10^{-3}$ within a generic grand unified theory (GUT). Therefore, since the curvature must be very small nowadays, in what follows it will be assumed that $\nu =0$ for all values of the curvature with the above expression assuming the simplest form:

\begin{equation}\label{LambdaF}
8\pi G\rho_v (t) \equiv \Lambda (t)=\Lambda_\infty + 3\left(\frac{H}{H_I} \right)^n \left(H^2+\frac{\kappa}{a^2} \right)\,.
\end{equation}
In the flat case,  the above expression is suggested for even powers of $n$ by the covariance of the effective action in quantum field theory in curved spacetimes\cite{LBS13,PLBS13}. The flat case form  has also been recently adopted\cite{LBS14} to explain the now observed entropy  produced during the transition from the early de Sitter state to the standard radiation phase. For $n=1$, $\Lambda_{\infty} = 0$ and $\kappa \neq 0$, it also reduces to the form adopted long ago by Lima and Trodden\cite{LimaTroddenPRD96}.
As we shall see next,  the above expression provides the minimal model for a complete cosmological history when curvature is considered. Note that the second term in the above expression drives the vacuum dynamics at very early times when the $H^{2}$ term is important. However, as long as  $H << H_I$ the constant $\Lambda_{\infty}$ term dominates with the model following the $\Lambda$CDM  evolution. 

\section{Spacetime Dynamics: The Complete Scenario}

Let us now discuss how the phenomenological expression (\ref{LambdaF}) for  $\Lambda (H, a)$ links the rapid dynamics of the primeval de Sitter vacuum state  with the standard radiation phase, and, finally, this era with the current vacuum-matter accelerating phase.  In what follows we assume that the perfect-fluid component obeys the equation of state (EoS):

\begin{equation}\label{EoS}
p=(\gamma-1)\rho\,,
\end{equation}
where  the ``adiabatic index'' $\gamma\in[1,2]$. 

Now, by inserting (\ref{LambdaF}) into the field equations (\ref{rho_T})-(\ref{P_T}) one obtains the general equation governing the evolution of the scale factor:

\begin{equation}\label{ddot_a}
a\ddot a + (\frac{3\tilde\gamma-2}{2})\dot a^2 + (\frac{3\tilde\gamma-2}{2})\kappa -\frac{\gamma}{2}\,\Lambda_\infty a^2=0\,,
\end{equation}
where the EoS (\ref{EoS})  was used to obtain the above expression and the effective $\tilde \gamma$ is defined by:
\begin{equation}
\tilde\gamma=\tilde\gamma(H)\equiv\gamma\left[1-\left(\frac{H}{H_I}\right)^n\right]\,.
\end{equation}

\begin{figure}[!ht]\centering
\includegraphics[width=8cm]{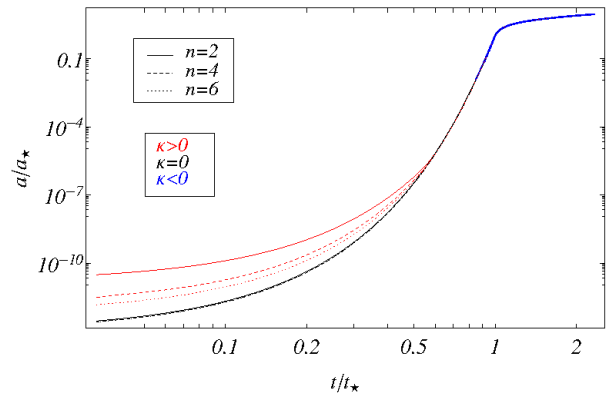}
\caption{\text{Evolution of the scale factor}. The figure shows the evolution of the normalized $a(t)$ from de Sitter to radiation phase for  different values of n and $\kappa$. Note that the hyperbolic case (blue line) has a minimal value $a_{min}$ fixed by the validity of the weak energy condition while for the closed we adopted the the Planck length in order to guarantee  the validity of the classical model. }\label{fig:hyper} .
\end{figure}

As it will be discussed next, the equation of motion (\ref{ddot_a}) describes a spacetime evolving from an early de Sitter phase to the present $\Lambda$CDM cosmology. It shows that  the same actor (decaying $\Lambda(t)$-term) can be responsible for the different accelerating phases. 

\subsection{From de Sitter to Radiation Phase}

As remarked before, at early times the $\Lambda_{\infty}$-term in the motion equation can be neglected. At this limit, the ratio between the fluid and vacuum energy density can be written as:
\begin{equation}\label{rho}
\frac{\rho}{\rho_v} = \left({\frac{H_I}{H}}\right)^n - 1\,, 
\end{equation}
and since $\rho_v$ is always positive, the fluid component will obey the weak energy condition, $\rho \geq 0  \Leftrightarrow H \leq H_I$. Therefore, for all values of $\kappa$, we see that $H_I$ is the maximum value physically allowed for the Hubble parameter. It is also interesting that for $H=H_I$ there is no fluid component. In particular,  the flat model starts from a pure de Sitter state with constant Hubble parameter, determined by the maximum vacuum energy density, $\rho_{vI} = 3H_I^{2}/8\pi G$.  When $H  \rightarrow H_I$ we see that  $\tilde\gamma  \rightarrow  0$ so that the equation \eqref{ddot_a} simplifies to:

\begin{equation}
a\ddot a - \dot a^2 - \kappa =0\,,
\end{equation}
which is exactly equation (\ref{EqSitter}) whose solutions assume the familiar de Sitter forms (\ref{desitter}) and represent the different classes of the de Sitter spacetimes (see section 2).  However,  it should be recalled that the Hubble parameter is not constant for a spatially curved de Sitter spacetime ($\kappa =\pm 1$). This result is confirmed by the expressions for the vacuum and matter energy densities at this limit: 
\begin{equation}
8\pi G\rho_v \equiv \Lambda=3\left(\frac{H}{H_I} \right)^n\left(H^2+\frac{\kappa}{a^2} \right)\xrightarrow{H\to H_I}3\left(H^2+\frac{\kappa}{a^2} \right) \equiv \rho_T\,,
\end{equation}
and
\begin{equation}
8\pi G\rho=3\left[1-\left(\frac{H}{H_I} \right)^n \right]\left(H^2+\frac{\kappa}{a^2} \right)\xrightarrow{H\to H_I}0\,,
\end{equation}
The basic result here is that the ansatz \eqref{ansatz} implies that the model starts from a pure de Sitter accelerating stage. 
However, since the $\Lambda$ term is a rapidly decreasing function with the expansion, one may expects that the universe ultimately it will be driven to a decelerating stage  becoming dominated by the non vacuum component (radiation for $\gamma = 4/3$). Naturally, such a transition must be directly connected with a solution to the so-called ``exit problem'' of inflation. 

\begin{figure}[!ht]\centering
\includegraphics[width=6.2cm]{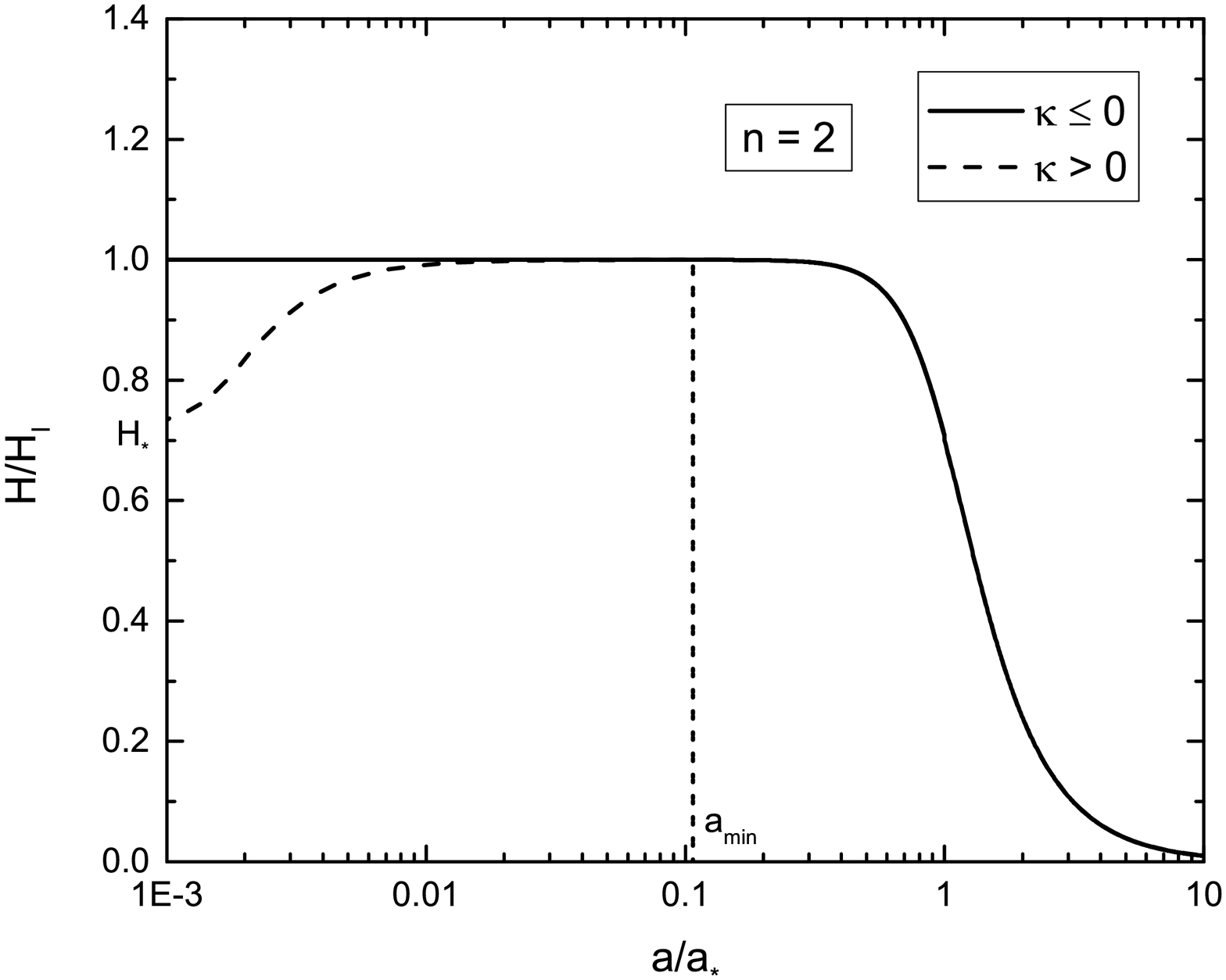}
\includegraphics[width=6.2cm]{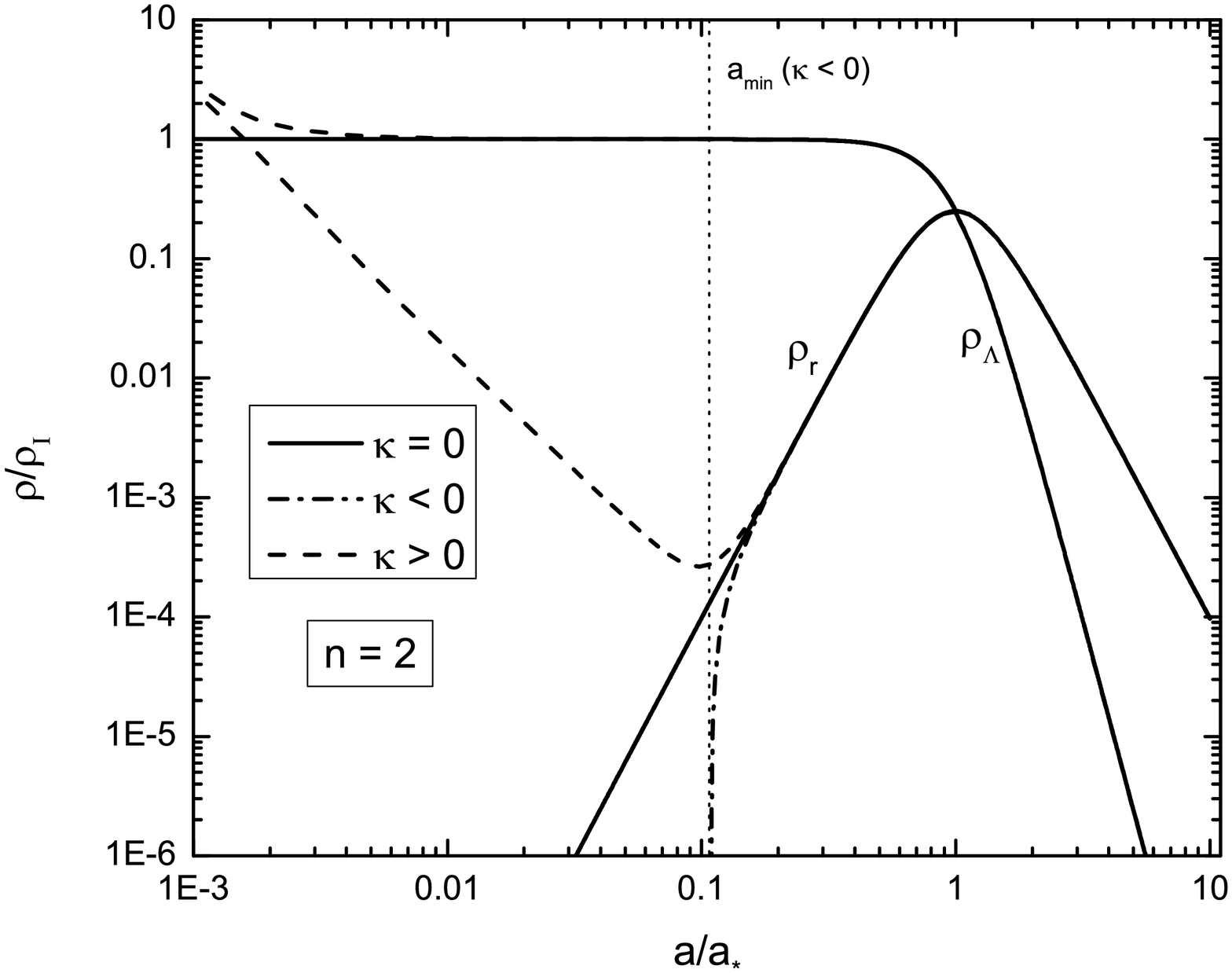}
\caption{Evolution of the Hubble parameter (left) and the associated energy densities (right) for the different values of $\kappa$.}\label{fig:hyper} .
\end{figure}
In order to study with more detail the early dynamics of the general model we must solve the field equations. However, except for the flat case,  it is not possible to obtain closed analytical forms for the Hubble parameter and the energy densities, and, therefore, numerical techniques are required. 



In figure 4,  we  display the results for the associated (normalized) Hubble parameters (left) and energy densities (right) for the case $n=2$, as a function of the normalized scale factor. Note that $H \leq H_I$ as qualitatively discussed before. However, the result is rigorously valid only for $\kappa = 1,0$. In the the hyperbolic case, the solution was truncated at $a=a_{min}$ by the weak energy condition. This is needed because otherwise $H$ may increase without limit and the fluid energy density becomes negative. The dashed line for the closed case show the influence of the curvature at early times. The value of $H$ diminishes in order to compensate for the increasing curvature contribution.  In the right plot the energy density was normalized by the critical density determined by the vacuum energy density of the flat case. We see the minimal value of the scale factor in the hyperbolic case when the fluid energy density goes to zero. Note that in the closed case the energy density of both components grow due to the curvature contribution even when the Hubble parameter diminishes. For all practical purposes, all models have the same evolution after $a=a_{min} \sim H_I^{-1}$.  

{\it When  the inflationary process ends}? So far, it was not possible to find a first integral  to the general equation of motion (\ref{ddot_a}). It can analytically be solved only in the flat case\cite{PLBS13}. However, it is possible to show that at early times the decelerating parameter $q(t)$ for all values of the curvature is given by:
\begin{equation}\label{Q}
q(t) = \frac{3\tilde\gamma-2}{2}\left(1 + \frac{\kappa}{a^{2}H^2}\right),
\end{equation}
and assuming that the vacuum decays in ultrarelativistic particles ($\gamma = 4/3$) one finds that the begin of the decelerating regime ($q=0$) will occur when the Hubble parameter becomes $H = H_I/2^{1/n}$, a value that is independent of the curvature parameter. In addition, from equation (\ref{rho}) we also see that  $q=0$ defines exactly the moment of equality vacuum-radiation energy densities ($\rho_v = \rho_{rad}$). Naturally,  due to the expansion the vacuum energy density becomes rapidly subdominant thereby starting the radiation domain. Therefore, the Universe as described here evolves continuously from inflation to the standard radiation era regardless the value of $\kappa$. For $H=H_I$ the radiation energy density is zero, but it increases continuously powered by the decaying vacuum component. Different from many inflaton-like formulations,  the inflationary process here is neither isentropic nor isothermal. In certain sense it resembles the so-called warm inflationary scenarios where the strong coupling of the inflaton field may alter both the slow roll conditions and the adiabatic evolution in such a way that the huge expansion is concomitant with the entropy production\cite{Berera95,BR03,ML99}. A detailed thermodynamic study about the generation of the radiation entropy in the flat case has recently been  discussed in the literature\cite{LBS14}.

\subsection{From radiation to vacuum-matter phase}

As one may expect, when the inflation ends ($q=0$) the vacuum component is still powerful enough to drain some energy (and entropy)  to the radiation component. However, after some time,  the increasing radiation entropy saturates in its final value thereby starting the adiabatic radiation phase (see Lima, Basilakos \& Sol\`a\cite{LBS14} for a detailed  analysis  in the flat case). At this moment, $H \ll H_I$ and the energy of the coupled part of the vacuum energy component has been fully consumed with the $\Lambda$-term also attaining its final value ($\Lambda_{\infty}$) with $\dot{\Lambda}=0$ (see Eq. \ref{LambdaF}). Hence, $\tilde \gamma = \gamma$ (the fluid component becomes separately conserved) and the dynamic equation \eqref{ddot_a}  takes the simple form:
\begin{equation}
a\ddot a+ (\frac{3\gamma-2}{2})\dot a^2 + (\frac{3\gamma-2}{2})\kappa- \frac{\gamma}{2}\Lambda_{\infty}a^2=0\,.
\end{equation}
The first integral of this equation can be written as:
\begin{equation}
{\dot{a}}^{2}= Aa^{2-3\gamma}-\kappa + \frac{\Lambda_\infty}{3}a^{2}\,.
\end{equation}
where A is an integration constant. For the sake of completeness, during these stages  of the universe (radiation and vacuum-matter phases), the vacuum and  fluid energy densities can be written as:
\begin{equation}
8\pi G\rho_v \equiv \Lambda=\Lambda_\infty\,,
\end{equation}
\begin{equation}
\rho=\rho_{\gamma0}(\frac{a_0}{a})^{3\gamma}\,,
\end{equation}
which corresponds for $\gamma=1$ to the current $\Lambda CDM$ model.

\section{Final Comments}

We have shown that the large class of spatially flat models describing a complete cosmological history\cite{PLBS13} occurring between two extreme  early and late time de Sitter phases  intercalated by two decelerating  stages  -  radiation and  cold dark matter phases - can be extended to any value of the curvature parameter ($\kappa = \pm 1,0$).  
The  early de Sitter stage is fully dominated by a phenomenological dynamical $\Lambda$-term, characterized by an arbitrary energy scale $H_I$ which can be adjusted to satisfy the CMB constraints. The proposed  phenomenological expression $\Lambda (H,a,\kappa)$  is responsible for the transition between the early de Sitter stages for the FLRW radiation era where the curvature is negligible. The main results derived here may be summarized in the following statements:

1) Regardless of the values assumed by the curvature parameter, the  decaying $\Lambda$-model discussed here  alliviates the cosmological constant and coincidence problems because the huge vacuum energy density that drives the early de Sitter phase is continuously transfered to the radiation component. At the end of the process ($H << H_I$),  the only remaining $\Lambda$-term is  the final constant value, $\Lambda_{\infty}$ (see Eq. (\ref{LambdaF})).

2) The instability of the early de Sitter phase is described by a nonadiabatic process and the generation of the radiation entropy is concomitant with inflation.  The transition from the  early de Sitter stage to the radiation phase is smooth and independent of the curvature parameter (see Eq.(\ref{Q})). However, it depends on the value of the power index $n$ appearing in the phenomenological decay law. The transition is faster for higher values of $n$. 

3) The models with $\kappa = 1,0$ are nonsingular and solves naturally the horizon and the ``graceful exit'' problems.  However, a preliminary numerical  treatment  suggest that the hyperbolic case does not solve properly the horizon problem. In this case,  the scale factor has a minimum value, $a=a_{min}$, which is required by the validity of the weak energy condition.

4) If the characteristic scale $H_I$ is adjusted to be of the order of the Planck energy, the ratio $\rho_I/\rho_{v0} \sim 10^{123}$ for all values of the curvature.  

We would like to stress that a more detailed analysis is needed to  obtain the amount of  the entropy produced in the general case.  The closed scenario deserves a special attention because it is endowed with very nice properties to describe the begin of the Universe (zero net charge, zero total angular momentum and energy, etc.),  and, probably, more important, the joint analyses involving complementary observations (WMAP and Planck) have show that the space parameter is still mildly compatible with  $\Omega_{T} > 1$. Naturally, if the cosmic history occurs between two extreme de Sitter phases,  the instability of the de Sitter spacetimes\cite{mottola} suggests that the Universe may be cyclic in the long run.  In other words,  the future  de Sitter phase can be the beginning of a `new cycle'.

\section*{Acknowledgments}

J.A.S.L. is partially supported by CNPq and FAPESP (Brazilian Research Agencies), E.L.D.P. and G.Z. are supported by  fellowships from CNPq and CAPES, respectively.

\end{document}